\documentclass[conference]{IEEEtran}

\usepackage[mathlines]{lineno} 

\newif\ifdraft
\draftfalse % set \drafttrue for internal drafts

\usepackage{cite}
\usepackage{amsmath,amssymb,amsfonts}
\usepackage{algorithm}
\usepackage{algpseudocode}
\usepackage{booktabs}
\usepackage{multirow}
\usepackage{colortbl}
\usepackage{graphicx}
\graphicspath{{./}{paper/}}
\usepackage{textcomp}
\usepackage{xcolor}
\usepackage{listings}
\usepackage{hyperref}
\usepackage{tabularx}
\usepackage{array}
\usepackage{siunitx}
\usepackage{tikz}
\usetikzlibrary{arrows.meta, positioning, decorations.pathreplacing, fit, backgrounds, matrix}
\usepackage[most]{tcolorbox}
\tcbuselibrary{listings,breakable}

\hypersetup{hidelinks}

\definecolor{promptbg}{RGB}{248,249,250}
\definecolor{promptframe}{RGB}{52,73,94}
\tcbset{
  promptbox/.style={
    enhanced, breakable,
    width=\columnwidth,
    colback=promptbg,
    colframe=promptframe,
    arc=2pt, outer arc=2pt,
    boxrule=0pt,
    before skip=1pt, after skip=1pt,
    left=3pt, right=3pt, top=1pt, bottom=1pt,
    fonttitle=\sffamily\bfseries\footnotesize,
    colbacktitle=promptframe,
    coltitle=white,
    titlerule=0pt,
    toptitle=1pt, bottomtitle=1pt,
  }
}

\newcommand{\code}[1]{{\small\ttfamily #1}}

\definecolor{codegreen}{rgb}{0.25,0.49,0.31}
\definecolor{codegray}{rgb}{0.50,0.50,0.50}
\definecolor{codepurple}{rgb}{0.44,0.12,0.49}
\definecolor{codeblue}{rgb}{0.13,0.29,0.53}
\definecolor{backcolour}{rgb}{0.97,0.97,0.97}
\definecolor{rulecolour}{rgb}{0.75,0.75,0.75}
\definecolor{diffgreen}{HTML}{E6FFEC}
\definecolor{diffred}{HTML}{FFEBE9}
\definecolor{diffgreenbar}{HTML}{1A7F37}
\definecolor{diffredbar}{HTML}{CF222E}
\definecolor{tblhead}{RGB}{223,231,242}   % column-header row
\definecolor{tblsub}{RGB}{238,242,248}    % section-divider rows
\definecolor{tblhl}{RGB}{210,226,248}     % highlighted result row
\definecolor{aznamebg}{RGB}{33,53,85}
\definecolor{aznameborder}{RGB}{22,37,58}

\newtcbox{\aznamebox}{
  on line,
  boxrule=0.5pt,
  colback=aznamebg,
  colframe=aznameborder,
  coltext=white,
  arc=1.2pt,
  boxsep=1pt,
  left=3pt,
  right=3pt,
  top=1pt,
  bottom=1pt,
  fontupper=\sffamily\bfseries\scriptsize
}

\ifdraft
  \newcommand{\az}[1]{\textcolor{red}{\aznamebox{Arastoo}~#1}}
  \newcommand{\mv}[1]{\textcolor{red}{\aznamebox{Marco}~#1}}
\else
  \newcommand{\az}[1]{}
  \newcommand{\mv}[1]{}
\fi
\ifdraft
  \newcommand{\gap}[1]{\textcolor{red}{\textbf{[GAP]}~\textit{#1}}}
\else
  \newcommand{\gap}[1]{}
\fi
\begin{document}

\title{AutoTrace: From Patches to Triggers via Agentic~Interprocedural Exploration}

\author{
\IEEEauthorblockN{Arastoo Zibaeirad}
\IEEEauthorblockA{University of North Carolina at Charlotte\\
Charlotte, NC, USA\\
azibaeir@charlotte.edu}
\and
\IEEEauthorblockN{Marco Vieira}
\IEEEauthorblockA{University of North Carolina at Charlotte\\
Charlotte, NC, USA\\
marco.vieira@charlotte.edu}
\and
\IEEEauthorblockN{Thomas Zimmermann}
\IEEEauthorblockA{University of California, Irvine\\
Irvine, CA, USA\\
tzimmer@uci.edu}
}

\maketitle

\begin{abstract}
Given a vulnerability-fixing commit, \emph{trigger localization} asks which specific
statement turns the vulnerable program state into a concrete unsafe operation.
This question is harder than binary vulnerability detection because the answer
demands interprocedural, causal reasoning: in a substantial fraction of real-world
CVEs the triggering statement lies several call layers outside the patched function,
beyond the reach of static rule sets and pattern-matching language models alike.
We present \emph{AutoTrace}, an agentic pipeline that localizes vulnerability
triggers by exploring a code property graph layer by layer, with LLM agents
deciding where to look next and deterministic admissibility gates deciding what
evidence is required before a trigger can be reported.
Agents never accept a trigger on their own authority; every reported trigger is
backed by explicit evidence drawn from the graph, so the pipeline covers both
intra- and interprocedural vulnerabilities without relying on ungrounded model
judgment.
On the full InterPVD benchmark, AutoTrace reaches 75.0\% VulnHit and 80.8\%
FuncHit, surpassing the prior state of the art on the same corpus.
Building on the same machinery, we construct \emph{SinkTrace-Bench}, a dataset
that exposes each vulnerability as a source-to-sink (S2S) causal chain from
attacker-controlled input through propagation to the dangerous operation, drawn
from matched vulnerable and patched program states. It comprises 1{,}542
verifier-confirmed, perfectly balanced vulnerable/safe samples whose label
fidelity we audit against expert annotations. Benchmarking frontier LLMs on it,
we find that even the strongest struggle to separate the matched pairs, exposing
the causal-reasoning gap that trigger localization targets. Artifact available at \url{https://github.com/Erroristotle/AutoTrace}.
\end{abstract}


\section{Introduction}
\label{sec:introduction}

\begin{figure*}[!t]
\centering
\includegraphics[width=0.92\textwidth]{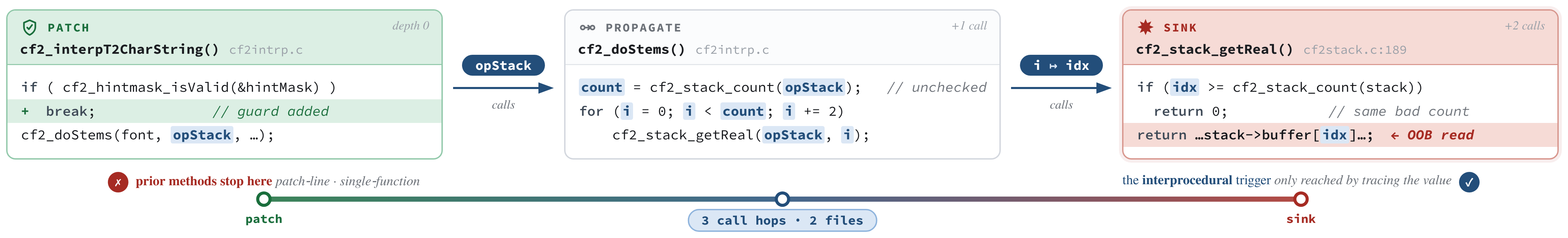}
\caption{Motivating example (CVE-2014-9659, CWE-125, FreeType). The patch adds a bounds guard in \code{cf2\_interpT2CharString}, yet the trigger (the \emph{sink} where the corrupted index is read, \code{stack->buffer[idx]}) lies three call layers away in another file. Patch-line and single-function methods point at the guard; only statement-level interprocedural localization reaches the sink.}
\label{fig:motivation}
\end{figure*}

Modern software-security work increasingly asks not only \emph{whether} a function is vulnerable but \emph{where} the vulnerability is exercised. Given a vulnerability-fixing commit, \emph{trigger localization} identifies the specific statement at which a vulnerable program state becomes a concrete unsafe operation such as a memory access, an allocation, or a dereference. Unlike binary detection, this demands causal reasoning across the call graph, because the triggering statement is frequently not the patched line and often lies in a different function from the patch. Prior empirical studies report that a substantial fraction of real-world C/C++ CVEs are interprocedural in exactly this sense, with patch and trigger separated by several call layers and sometimes several files~\cite{li2024effectiveness,sejfia2024toward}.

This precision is what makes the task worth solving, because the coarser artifacts a security engineer is usually handed each answers a weaker question: a function-level label only says a defect lies somewhere in a long body, a patch hunk often marks a guard inserted upstream rather than the unsafe operation, and a S2S slice delimits a region without naming the operative statement. Only the trigger statement pins the exact operation, on a specific operand and under specific guards, which is what patch review, exploitability triage, test construction, and causal labeling of training data actually consume. Getting it wrong is costly: a false trigger sends a reviewer to the wrong site and leaves the real defect live, and a dataset built from syntactic triggers teaches models to imitate changed-code patterns instead of causal structure. These failure modes shape our design: AutoTrace reports a trigger only when non-bypassable evidence supports it, and SinkTrace-Bench labels causal structure rather than patch syntax.

Existing approaches struggle on this class of vulnerability: rule-based tools miss project-specific wrappers and macro-expanded helpers, single-function line rankers cannot follow taint across a call boundary, and VulTrigger~\cite{li2024effectiveness}, the only prior system targeting trigger localization from a fixing commit, classifies interprocedural slices by syntactic rule matching that conflates co-occurrence with causal connection (\S\ref{sec:related-work}). We are careful about what Figure~\ref{fig:motivation} establishes: it shows that single-function analysis is insufficient, not that precise interprocedural analysis is impossible. In principle a sufficiently precise analysis could resolve it; the practical difficulty, developed in \S\ref{subsec:difficulty}, is that the alias, field, dispatch, macro, and control-dependence precision such an analysis needs is rarely available on real C/C++, and that is where LLM-guided exploration earns its place.

We present \textbf{AutoTrace}, an agentic pipeline that separates two concerns. LLM agents act as semantic reasoners that interpret code, extract the variables a vulnerability centers on, and decide where to look next, while deterministic admissibility gates over a code property graph~\cite{yamaguchi2014modeling,joern} fix the hard conditions a trigger must satisfy before it is reported, so no trigger rests on unverified model judgment. On the full InterPVD benchmark, AutoTrace reaches 75.0\% VulnHit and 80.8\% FuncHit, outperforming VulTrigger's 69.8\% on the same corpus under the same line-level inclusion metric. On the cohort it shares with a retrieval-augmented LLM baseline, AutoTrace localizes the exact trigger nearly twice as often (\S\ref{subsec:rq1}); the baseline's higher function-level reach is a by-product of never abstaining, whereas AutoTrace's CPG-grounded gate reports only what it can certify.

The same machinery yields \textbf{SinkTrace-Bench}, a dataset whose samples expose each vulnerability as a S2S chain from an attacker-controlled source through propagation to the dangerous sink, drawn from paired vulnerable and patched program states. Prior vulnerability datasets label \emph{what} is vulnerable but not \emph{why}, so models trained on them learn surface patterns when the task demands causal reasoning; SinkTrace-Bench instead annotates causal structure. Because its labels are produced by AutoTrace's verifier rather than by independent experts, we describe them as verifier-confirmed rather than expert-validated, and we audit their agreement with expert ground truth in \S\ref{subsec:rq3}. Every sample belongs to a matched vulnerable/safe pair drawn from the same source-to-sink chain and differing only by the patch, so the release is a balanced set of 771 such pairs (1{,}542 samples) across 16 CWE classes.

\noindent\textbf{Contributions.}
\begin{itemize}\setlength\itemsep{1pt}
    \item \textbf{AutoTrace}, an agentic pipeline whose defining property is acceptance discipline: LLM agents propose candidate triggers and non-bypassable deterministic gates over a code property graph decide, so no reported trigger rests on unverified model judgment. Its measured advantage over simple LLM and retrieval baselines is interprocedural trigger localization (VulnHit).
    \item \textbf{SinkTrace-Bench}, 771 matched vulnerable/safe pairs (1{,}542 balanced verifier-confirmed S2S samples) spanning 16 CWE classes, with the causal chain as a first-class label, released together with an audit of label fidelity against expert annotations.
\end{itemize}

\S\ref{sec:related-work} surveys related work, \S\ref{sec:methodology} defines the task and the pipeline, \S\ref{sec:evaluation} reports the experiments, and \S\ref{sec:discussion} and \S\ref{sec:conclusion} discuss limitations and conclude.

\section{Related Work}
\label{sec:related-work}

Learning-based vulnerability analysis has moved along two axes: finer-grained \emph{outputs}, from function classification to line ranking to statement-level trigger identification, and richer \emph{evidence}, from token sequences to program slices to code property graphs (CPGs)~\cite{yamaguchi2014modeling,joern} and, recently, LLM agents that consult program-analysis tools on demand. AutoTrace sits at the end of both: its output is a specific trigger statement backed by an interprocedural causal chain, and its evidence comes from an agent whose every acceptance is gated by a deterministic CPG-based verifier.

\paragraph{Localization and hybrid LLM/static analysis.}
Slice-based detectors showed that structured program context beats raw tokens for classification, and a line of work refines the slice fed to the model~\cite{li2021sysevr,wu2023learning,salimi2022vulslicer,machtle2025trace,lekssays2025llmxcpg,chen2024utilizing}, while intra-function line rankers~\cite{fu2022linevul,ding2022velvet,tang2024mrc} score statements within a known-vulnerable function. Neither family reaches interprocedural triggers: line rankers are bounded by the function under analysis, and VulTrigger~\cite{li2024effectiveness}, the only prior system to target trigger localization from a VFC, pairs critical-variable slices with CWE keyword matching, so it cannot enumerate project-specific wrappers and collapses syntactic co-occurrence into causal connection. Hybrid LLM/static approaches such as IRIS~\cite{li2024iris} and CPGHunter~\cite{hou2026cpghunter} use LLMs to augment static analysis but share the objective of classical taint analysis, whether \emph{any} S2S path exists, rather than identifying the statement where vulnerable state becomes an unsafe operation.

\paragraph{Why classical and dynamic analyses do not subsume the task.}
A fair positioning must reach past line rankers to the interprocedural analyses that already track sources, sinks, call chains, aliases, and guards. AutoTrace's own non-agentic core, CPG slicing plus deterministic gates, is itself such an analysis, so the open question is what the agent adds on top, which we isolate in~\S\ref{subsec:rq2}. Classical IFDS/IDE dataflow answers ``does any source reach any sink'' against a fixed sink specification, whereas trigger localization supplies a patch-derived source, needs an open project-specific sink vocabulary, and must return the single root-cause statement. Demand-driven slicing and dependence graphs yield a slice, not the operative statement, and cannot separate a causal statement from one that merely co-occurs with the tracked value. Query engines such as CodeQL require a hand-authored query per CWE and sink family; the 72-CVE CodeQL point in Table~\ref{tab:rq1_main} (8.6\%) reflects that brittleness. Dynamic techniques (symbolic execution, taint, fuzzing) change the setting rather than strengthen the static baseline: they need a build, a harness, and concrete inputs, are not patch-conditioned, and report a crash site, a consequence rather than the root-cause trigger. None is weak in general; each is mismatched to this task, and the precision it would need on real C/C++ is exactly what degrades in practice (\S\ref{subsec:difficulty}).

\paragraph{Agentic code analysis.}
The closest neighbor is RepoAudit~\cite{guo2025repoaudit}, an autonomous LLM agent for repository-level bug detection through demand-driven call-graph exploration. AutoTrace adopts the same agent-over-static-evidence paradigm but differs on three axes. It is VFC-conditioned, recovering the causal chain a patch neutralizes rather than performing open-world discovery. It separates agent proposals from deterministic acceptance through the CPG gates, so no report rests on unverified judgment, whereas RepoAudit's final filter is LLM-judged path feasibility. And it anchors its exploration budget to the first verified trigger rather than a fixed call-context cap, spending effort on deepening a chain.

\section{Methodology}
\label{sec:methodology}

\subsection*{Overview}

\begin{figure*}[!t]
\centering
\includegraphics[width=1.0\textwidth]{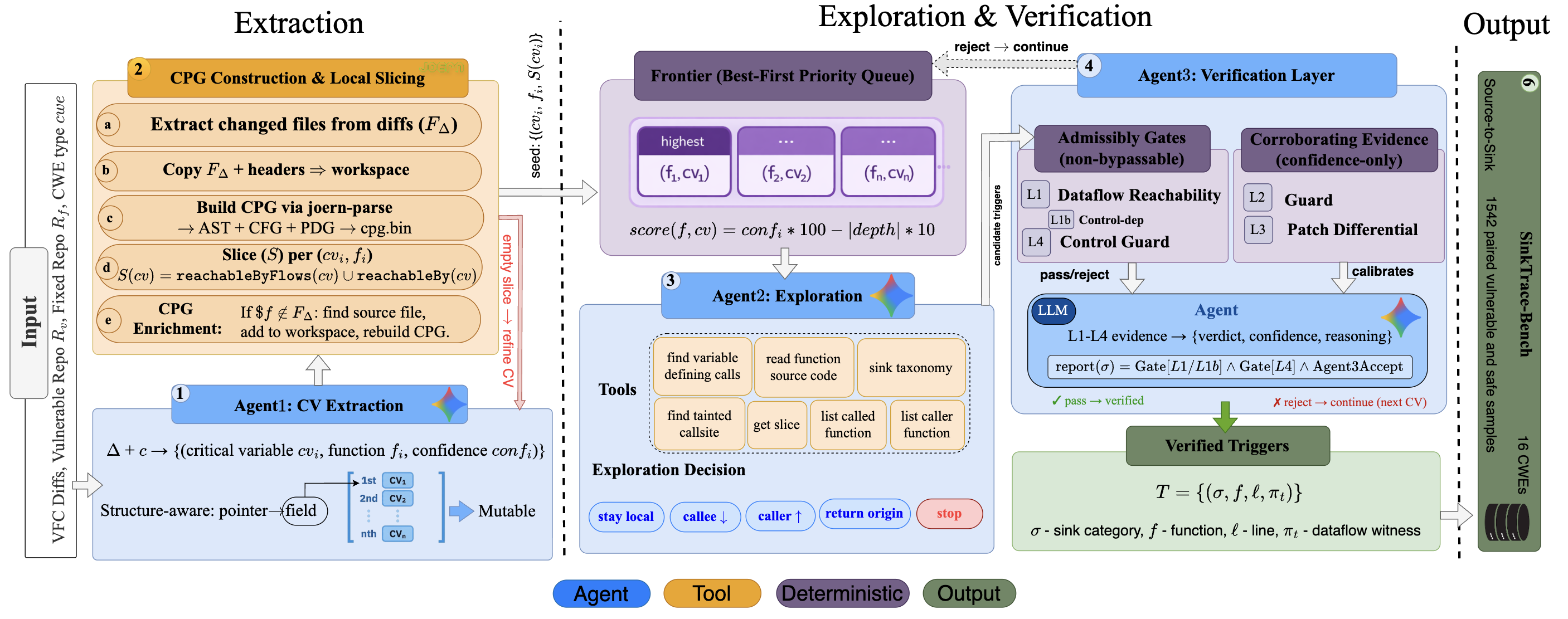}
\caption{AutoTrace pipeline: CPG construction, best-first frontier search, deterministic gate verification, and SinkTrace-Bench construction.}
\label{fig:framework}
\end{figure*}

AutoTrace takes a vulnerability-fixing commit (VFC) with its pre- and post-fix repository states and returns the trigger statements of \S\ref{subsec:problem-definition}, each a single statement~$\sigma$ certified by, but distinct from, a witness path, including triggers several call layers outside the patched function. Two principles govern the design. First, LLM agents act as semantic reasoners that interpret code and decide where the search moves next, but never bypass the hard admissibility checks: every candidate must survive deterministic gates that test it against explicit code-property-graph (CPG) evidence, which keeps the model's flexibility with project-specific wrappers while the verifier keeps outputs grounded. Second, interprocedural analysis proceeds through \emph{repeated local slicing} rather than one whole-program slice, with each function sliced independently and evidence stitched across call edges, which bounds both analysis and prompt context to a single function however deep the chain. Figure~\ref{fig:framework} summarizes the five stages, from critical-variable extraction (Stage~1) through the patch-scoped CPG and slicing (Stage~2), agentic best-first exploration (Stage~3), and verification gates with a synthesizing agent (Stage~4), to SinkTrace-Bench construction (Stage~5).

\subsection{Problem Formulation}
\label{subsec:problem-definition}

We formulate trigger localization as a solver-agnostic computational problem.
All five equations below are definitional; how any system approximates them is its own concern.

\paragraph{Vulnerability instance.}
A \emph{vulnerability instance} is the five-tuple
\begin{equation}
  \mathcal{V} \;=\; \bigl(P^{-},\;P^{+},\;\Delta,\;C_{\Delta},\;w\bigr),
  \label{eq:instance}
\end{equation}
where $P^{-}$ is the vulnerable program snapshot, $P^{+}$ is its fixed counterpart,
$\Delta$~is the vulnerability-fixing commit, $w$~is the CWE class, and
$C_{\Delta}$~is the set of \emph{patch-anchored critical values} — variables, fields,
parameters, or expressions whose value, bounds, lifetime, or guard is constrained
by~$\Delta$.
We treat $C_{\Delta}$ as given with the instance; Stage~1
(\S\ref{subsec:cv-extraction}) describes how the pipeline infers it autonomously
from the diff.

\paragraph{Program dependence graph.}
We model $P^{-}$ as the directed multigraph
\begin{equation}
  D(P^{-}) \;=\; \bigl\langle\, S,\;
    E_{\mathrm{dd}} \cup E_{\mathrm{cd}} \cup E_{\mathrm{al}} \cup E_{\mathrm{call}}
  \,\bigr\rangle,
  \label{eq:pdg}
\end{equation}
over a statement set $S$, each node annotated with its enclosing function
$\mathrm{fn}(s)$ and source location $\mathrm{loc}(s)$.
The four edge types carry \emph{data dependence} ($E_{\mathrm{dd}}$),
\emph{control dependence} ($E_{\mathrm{cd}}$),
\emph{alias/field reachability} ($E_{\mathrm{al}}$), and
\emph{call/return binding} ($E_{\mathrm{call}}$).
Equation~\eqref{eq:pdg} abstracts the program's full semantics;
our implementation realizes it via a code property graph
(\S\ref{subsec:slicing}).
A critical value $c\in C_{\Delta}$ \emph{causally influences} statement~$\sigma$,
written $c\rightsquigarrow\sigma$, when there exists a path over
$E_{\mathrm{dd}}\cup E_{\mathrm{cd}}\cup E_{\mathrm{al}}\cup E_{\mathrm{call}}$
from a use of~$c$ to~$\sigma$ that preserves the tracked value, guard, or condition at
every step.
This closure of typed dependences is the \emph{only} criterion for causal influence;
patch-differential reachability and guard absence are admissibility conditions, not
part of the definition.

\paragraph{Trigger statement.}
A statement $\sigma\in S$ is a \emph{trigger} if and only if
\begin{equation}
  \mathrm{Sink}_{w}(\sigma)
  \;\wedge\;
  \exists\,c\in C_{\Delta}:\; c\rightsquigarrow\sigma.
  \label{eq:trigger}
\end{equation}
$\mathrm{Sink}_{w}(\sigma)$ asserts that $\sigma$ dispatches the
dangerous-operation class associated with CWE~$w$ on an operand derived from
$C_{\Delta}$; Table~\ref{tab:sinktypes} fixes $\mathrm{Sink}_{w}$ for each CWE family.
Equation~\eqref{eq:trigger} admits only \emph{sink-typed} statements with a confirmed
causal path: propagation assignments and branch predicates may explain reachability
but are not triggers, and a downstream crash site is a consequence rather than the
root cause.
Notably, the realizing operation need not be the syntactic memory access; when a
corrupted bound induces the vulnerability, the trigger is the size computation that
establishes the unsafe extent, even though the fault materializes at a later write.

\begin{table}[t]
\centering\footnotesize
\setlength{\tabcolsep}{4pt}
\renewcommand{\arraystretch}{1.0}
\caption{Trigger predicate $\mathrm{Sink}_{w}(\sigma)$ per CWE family.}
\label{tab:sinktypes}
\begin{tabularx}{\columnwidth}{@{} l >{\raggedright\arraybackslash}X @{}}
\toprule
\rowcolor{tblhead}
\textbf{CWE family} & \textbf{Dangerous operation at $\sigma$} \\
\midrule
Memory {\tiny(CWE-119/120/125/787/788)}       & Read/write at a computed offset        \\
Integer/size {\tiny(CWE-189/190/191)}         & Arithmetic bounding an access or alloc \\
Null deref.\ {\tiny(CWE-476)}                 & Dereference of a possibly-null pointer \\
Use-after-free {\tiny(CWE-416/415)}           & Use or release of a freed object       \\
Division {\tiny(CWE-369)}                     & Divide/modulo by a possibly-zero value \\
Infinite loop {\tiny(CWE-835)}                & Loop-bound or termination-condition use\\
Resource leak {\tiny(CWE-401/772)}            & Last reference lost without release    \\
\bottomrule
\end{tabularx}
\end{table}

\paragraph{Witness path.}
A trigger $\sigma$ satisfying~\eqref{eq:trigger} is \emph{reportable} only when
accompanied by a concrete realization of $c\rightsquigarrow\sigma$.
A \emph{witness} is the finite alternating sequence
\begin{equation}
  \pi \;=\; \langle\, n_0,\; e_1,\; n_1,\; \ldots,\; e_k,\; n_k \,\rangle,
  \label{eq:witness}
\end{equation}
where $n_0$ uses some $c\in C_{\Delta}$, $n_k=\sigma$, and each $e_i\in E$
is a real dependence edge of $D(P^{-})$ that preserves the tracked value,
guard, or condition.
The localized object is always the single statement $n_k=\sigma$; the path~$\pi$
in~\eqref{eq:witness} is evidence, not the output.
A witness is \emph{interprocedural} when $\pi$ contains at least one edge from
$E_{\mathrm{call}}$.

\paragraph{Localization problem.}
Given $\mathcal{V}$ from~\eqref{eq:instance} and $D(P^{-})$
from~\eqref{eq:pdg}, the task is to output the set of admissible pairs
\begin{equation}
\begin{split}
  \mathcal{T}^{\star} = \bigl\{\,(\sigma,\pi) :\;&
    \mathrm{Sink}_{w}(\sigma) \;\wedge\;
    \exists\,c\!\in\!C_{\Delta}.\; c\rightsquigarrow\sigma \\
    &\wedge\; \pi\ \text{admissible for}\ (\sigma,c)
  \,\bigr\},
\end{split}
  \label{eq:output}
\end{equation}
each trigger $\sigma$ paired with a certifying witness $\pi$.
The projections $\mathrm{fn}(\sigma)$ (FuncHit) and $\mathrm{loc}(\sigma)$
(VulnHit, Within$\pm$5) yield coarser evaluation views (\S\ref{subsec:setup})
but are not independent outputs.

\paragraph{Admissibility.}
A candidate $(\sigma,\pi)$ as in~\eqref{eq:output} is \emph{admissible} when its
witness is justified by $P^{-}$, $P^{+}$, and $\Delta$ alone.
Five conditions must hold jointly:
\begin{enumerate}
  \item[(i)]   \textit{Patch anchoring.} $\pi$ originates at a value in $C_{\Delta}$
               or one provably equivalent to it under $\Delta$.
  \item[(ii)]  \textit{Trigger typing.} $\mathrm{Sink}_{w}(\sigma)$ holds per
               \eqref{eq:trigger} and Table~\ref{tab:sinktypes}.
  \item[(iii)] \textit{Causal reachability.} Every edge of~$\pi$ is a real dependence
               in $D(P^{-})$, so $c\rightsquigarrow\sigma$ genuinely holds rather than
               $\sigma$ merely co-occurring with~$c$.
  \item[(iv)]  \textit{Call-context realizability.} Call/return edges in~$\pi$ form a
               matched, realizable sequence.
  \item[(v)]   \textit{Patch relevance.} Comparing $P^{-}$ with $P^{+}$ confirms
               that $\Delta$ alters the reaching value, its guard, or the operation
               at~$\sigma$ to neutralize the unsafe relation.
\end{enumerate}
These conditions are intrinsic to $\mathcal{V}$ and solver-agnostic; how AutoTrace's
verifier enforces them is described in \S\ref{subsec:verification}.

\paragraph{Evaluation metrics.}
A solver's output $\hat{\mathcal{T}}$ is compared to a ground-truth set
$\mathcal{G}$ at three granularities.
\emph{VulnHit}: a predicted $\sigma$ matches $\sigma^{\star}\in\mathcal{G}$ when
both denote the same operation after macro canonicalization (statement level).
\emph{FuncHit}: $\mathrm{fn}(\sigma)=\mathrm{fn}(\sigma^{\star})$ (function level).
\emph{Within}$\pm5$: $|\mathrm{loc}(\sigma)-\mathrm{loc}(\sigma^{\star})|\le 5$
(line proximity).
An admissible report with no match in $\mathcal{G}$ is internally valid but
incorrect; a report backed by an inadmissible witness is an unsupported guess.

\subsection{Sources of Difficulty}
\label{subsec:difficulty}

The motivating example (Fig.~\ref{fig:motivation}) is easily misread as proof that interprocedural triggers \emph{require} an LLM. They do not, and we keep three claims distinct. \emph{Claim~1}, established: single-function line ranking cannot reach interprocedural triggers, since a ranker bounded by one function body cannot follow a value across a call edge to where it is consumed. \emph{Claim~2}, the operative difficulty: a sound but imprecise interprocedural analysis can fail in practice, and this rather than undecidability is what makes the task hard. A sufficiently precise analysis could in principle resolve Fig.~\ref{fig:motivation}, but the required precision is rarely available on real C/C++, where the critical value must survive pointer aliasing, field-sensitive dataflow, indirect dispatch, macro expansion, project-specific sink wrappers, and guard-conditional reachability. Each is a point where a classical pipeline silently loses the chain, and our results show the cost: the verifier refuses candidates exactly when Joern cannot establish a path through aliasing, indirect dispatch, or incomplete edges (\S\ref{subsec:rq2}). \emph{Claim~3}, which we do \emph{not} make: agentic exploration is not the only way to solve these cases, only an effective practical response to Claim~2, supplying the semantic judgment of which wrapper is a sink and which alias to follow where precise static analysis is unavailable, while every accept or reject is delegated to the deterministic gates. A more precise static analyzer that closed the Claim-2 gaps would be an equally valid solver under our formulation.

A conceptual difficulty compounds these. The trigger is not always the syntactic dangerous operation: when a corrupted bound induces the vulnerability, the root-cause trigger is the size computation, not the later write, so a localizer that targets the root cause and a benchmark that annotates the consequence disagree on a causally correct answer. We treat the resulting mismatch as a measured bias (\S\ref{subsec:threats}) rather than a system error.

\subsection{Threat Model}
\label{subsec:threat-model}

AutoTrace places LLM agents in a loop that reads untrusted source code and invokes external tools, so its threat model covers both the vulnerability under analysis and the integrity of the pipeline. The \emph{adversary} is the author of the repository content, and we assume control of all untrusted inputs: source code, comments, identifiers, macros, headers, repository layout, and the commit diff, including content crafted to steer a language model such as identifiers that read like instructions or misleading wrapper names. The adversary cannot modify Joern, the deterministic gates, the harness, or the ground-truth annotations. The protected \emph{asset} is the integrity of the trigger report, and transitively the SinkTrace-Bench labels derived from it. The \emph{trusted computing base} is the Joern CPG engine, the Admissibility Gates, and the harness that invokes them; the LLM backbone sits outside it and is treated as an untrusted reasoner. \emph{In scope} is adversarial influence on the LLM, including prompt injection through crafted comments or identifiers and over-eager localization; the structural defense is acceptance discipline, since a candidate enters the output only when its causal connection to a critical variable is backed by explicit CPG evidence, so a successful injection can at most yield a rejected or unexplored candidate, never an unsupported reported trigger. \emph{Out of scope} are attacks on the analysis substrate, which we treat as integrity limitations rather than defended threats, and a compromised Joern, harness, or backbone.

We are precise about what trusting the CPG buys: AutoTrace's guarantees are relative to the CPG, a sound but incomplete approximation of program semantics rather than a claim about the program, so wherever we write ``machine-checkable'' we mean checkable against this substrate and a passed gate attests reachability \emph{in the graph}, which may diverge from reachability in the program. AutoTrace bounds this exposure through the differential check, the control-dependence fallback, and a conservatism that prefers abstention over an unsupported deep guess; \S\ref{subsec:threats} treats the residual incompleteness as the principal integrity limit of a CPG-grounded verifier.

\subsection{Stage~1: Critical-Variable Extraction}
\label{subsec:cv-extraction}

The first stage identifies the \emph{critical variables}~(CVs), the named quantities whose misuse constitutes or propagates the vulnerability: an identifier (local, parameter, or field expression) that the patch introduces, removes, constrains, or newly protects, such as a length field the fix begins to bounds-check or a pointer whose lifetime it manages. An extraction agent reads the VFC diff with surrounding context and returns a structured list of CV candidates, each with the variable, its patch-facing location, a rationale, and a \emph{predicted} CWE family; the CWE is inferred by the agent rather than supplied, and is refined once a trigger's sink category is verified, so the pipeline never leans on a ground-truth label. Because downstream stages analyze CVs independently, one commit can yield multiple verified traces.

A patch-facing identifier is not always the carrier the verifier ultimately tracks, since the patch may name a struct field while the vulnerability flows through a local alias several statements later. AutoTrace therefore binds critical variables in a graduated fashion: it first attempts the exact field or parameter named in the patch and falls back to related carriers, such as the enclosing object or a local alias, only when the direct binding yields an empty slice, which keeps the search robust to aliasing without letting the agent invent unrelated carriers.

\subsection{Stage~2: Patch-Scoped CPG and Local Slicing}
\label{subsec:slicing}

\paragraph{Patch-scoped workspace.}
Rather than building a whole-repository CPG up front, AutoTrace assembles a \emph{patch-scoped workspace}: starting from the changed files, it grows on demand, adding callers, callees, and headers only when interprocedural exploration reaches them. Joern builds a CPG over this workspace, which keeps the analysis context small and anchors every traversal to the patch.

\paragraph{Repeated local slicing.}
Within each function AutoTrace slices locally for the current work item, recovering the statements that define, constrain, or alias the tracked value by backward slicing and exposing where it flows next by forward reachability. When the search crosses a call edge it re-slices locally in the target function and stitches evidence across the handoff rather than extending one interprocedural slice, which keeps each prompt focused on a single function instead of a whole-program excerpt. The slicing layer is state-aware: it analyzes the vulnerable commit to recover candidates and reuses the fixed commit for differential verification and dataset generation.

\subsection{Agent/Verifier Interface}
\label{subsec:agent-verifier-interface}

AutoTrace uses \emph{prompt-defined agents}, role-specific wrappers around one configured backbone rather than separately trained models. The only state passed between agents is typed JSON validated against the schemas below; free-form reasoning is logged for audit but never consumed as a decision. The interface separates \emph{candidate generation} from \emph{trigger certification}: Agent~2 may propose a sink, but only the verifier can certify it, and the verifier is itself wrapped by non-bypassable graph gates. LLM judgment is thus confined to proposing candidate locations and ranking among already-admissible ones, plus the single leak-family exception; no LLM verdict can place a gate-failing candidate into the output, so every reported trigger is attributable to graph evidence first and model judgment only as a filter.

\paragraph{Parsing and retry discipline.}
All agents use structured-output calls with conservative parsing: Markdown fences stripped, Pydantic validation, up to three retries with the schema and error appended. Hard gates remain the final authority: no retry or reflection loop can admit a candidate that fails L1/1b or L4. Agent~1 makes one broad-retry attempt on no usable variables; Agent~2 runs at most four turns of four tool calls with one reflection retry only when the slice exceeds five nodes and no sink has been found; Agent~3 makes at most three turns of two tool calls on ambiguous packets only, and cannot extend the loop to manufacture acceptance. All agents decode with the Gemini-3-Flash backbone at temperature~$1.0$, top-$p$~$0.95$, and a $4{,}096$-token cap (hop selection at~$0.2$); RQ3 (\S\ref{subsec:rq3}) decodes every model greedily at temperature~$0.0$.

\subsection{Stage~3: Agentic Interprocedural Exploration}
\label{subsec:frontier-search}

\paragraph{Verifier-gated best-first exploration.}
AutoTrace is interprocedural in reach but intraprocedural in each step: every turn is confined to one function, and the interprocedural chain is assembled by the agent's navigation decisions. We frame the search as an optimization problem, to find a high-confidence, shallow call-graph location at which the verifier accepts a candidate with confidence at least a pinned threshold~$\tau$, which turns open-ended traversal into targeted evidence-seeking. Once a critical variable is bound, the search is a best-first walk over a priority queue, the \emph{frontier}, of work items $(f,v)$ seeded from Stage~1. At each iteration the highest-priority item is dequeued, the function is sliced locally, and the agent decides whether a trigger is present and which callees, callers, or return-origin producers to enqueue. Each item carries a priority
\begin{equation}
\text{score}(f, v) = \underbrace{p_{\text{agent}} \cdot 100}_{\text{semantic relevance}} - \underbrace{|d| \cdot 10}_{\text{depth penalty}},
\end{equation}
where $p_{\text{agent}}\in[0,1]$ is the agent's confidence that $f$ merits exploration and $|d|$ is the call-graph distance from the patched function; the depth penalty biases toward shallow evidence at equal relevance. A visited set keyed on $(f,v)$ suppresses re-enqueuing, and an alias memory of every identifier the taint has entered and a remaining budget are carried across iterations so the search converges. Two further memories steer exploration without reaching the verifier: an \emph{episodic memory} that replays, within a run, sinks the gates already refused so they are not re-proposed, and a \emph{long-term memory} of cross-CVE priors by CWE and project (sink categories, expansion directions), category-level statistics that never bypass the gates. RQ2 (\S\ref{subsec:rq2}) ablates both.

The visited set bounds total work to $O(B)$ slices and agent turns; per-step cost scales with one function body, not the repository, so traversal is linear in budget and independent of program size.

\paragraph{A worked example.}
Consider the motivating CVE (Fig.~\ref{fig:motivation}, CVE-2014-9659). Stage~1 binds a stack-index value constrained by the patch in \code{interpT2CharString}. Iteration~1 slices that function, finds no local sink, and descends into \code{doStems}; iteration~2 follows the value into \code{stack\_getReal}; iteration~3 reaches line~189, the indexed read \code{stack->buffer[idx]}, where the verifier confirms the path \code{interpT2CharString}\,$\rightarrow$\,\code{doStems}\,$\rightarrow$\,\code{stack\_getReal} and accepts. The trigger lies three call layers from the patch, yet each step examined only one function's slice.

\paragraph{Confidence-triggered early termination.}
Each turn yields zero or more candidates that pass through the verifier of \S\ref{subsec:verification}; those failing a hard gate are discarded and do not affect termination. The search for a critical variable stops once a verified candidate has confidence $p_{\text{cand}}\geq\tau$ (default $0.95$) and the frontier holds no shallower unexplored item, with a two-level lookahead that preserves alternate high-confidence sinks for multi-sink CVEs. Two properties make this safe. Soundness is independent of the stopping rule: hard admissibility is owned by the deterministic gates, so stopping only changes which candidates are \emph{considered}, never whether one bypasses the evidence requirements. And because the depth-penalized order never stops while a shallower item remains, budget is spent deepening a concrete causal chain rather than on undirected discovery, unlike RepoAudit's fixed $K{=}4$ call-context cap that bounds exploration before anything is verified.

\paragraph{Tool interface.}
The exploration agent interacts with the CPG through a small library of semantic tools rather than raw graph queries. Each tool exposes one vulnerability-semantics question, such as ``does this value flow to this statement?'' or ``which calls take this value as an argument?'', and its deterministic implementation handles the correct CPG traversal. This keeps the agent's reasoning above the level of individual graph edges and prevents ad-hoc queries that silently diverge from Joern's semantics. Two auxiliary knowledge tools return CWE-specific sink categories and a short procedural workflow on demand. The full tool set and the underlying Joern queries are listed in Appendix~\ref{app:repro}.

\paragraph{Navigation decisions across call boundaries.}
At each work item the agent chooses one of four local moves (Fig.~\ref{fig:framework}): \emph{stay}, \emph{descend} into a callee, \emph{ascend} to a caller, or \emph{trace} a return-origin producer. The hard part is carrying the taint label across each scope change, mapping argument to parameter on descent and back on ascent; all tainted identifiers enter a per-sample alias memory, so the verifier can confirm a complete path even when no single name spans the chain. The agent also distinguishes \emph{leaf} calls that perform the dangerous action from \emph{routing} calls that merely forward the value, so the search neither stops at a forwarder nor descends uselessly into a leaf.

For each work item the agent returns a structured decision with candidate triggers, evidence citations, and expansion targets; if a turn yields no verified trigger and no useful expansion, it issues one reflection-based retry with explicit failure context. Verified triggers are later materialized as the S2S exemplars of Stage~5 (\S\ref{subsec:sinktrace-construction}).

\subsection{Stage~4: Multi-Layer Verification}
\label{subsec:verification}

Every candidate trigger must pass a multi-layer verification stage before it enters the output. The design rests on one principle: a report must be backed by at least one piece of \emph{necessary} evidence, and may be calibrated by \emph{corroborating} evidence that changes how much it is trusted but not whether it is admitted. The four layers split into two classes. \textbf{Admissibility Gates} (Layers~1 and~4) are hard rejections that no other evidence can override; \textbf{Corroborating Evidence} (Layers~2 and~3) modulates confidence but never vetoes. Which layers apply is set by the CWE class, and with the single leak-family exception noted in Threats to Validity the gates always apply.

\paragraph{Layer~1: Dataflow (gate).}
The primary gate requires an explicit dataflow path from the tracked critical variable~$v$ to the candidate~$\sigma$, $\mathtt{HasFlow}(v\to\sigma)=\mathit{true}$, paired with a concrete \code{FlowPaths} witness that is serialized into the output. When direct dataflow is absent, \emph{Layer~1b} recovers control-dependent triggers by checking whether $v$ participates in a loop bound or enclosing condition, whether $\sigma$ is control-dependent on that condition, and whether the unsafe access sits in the guarded region. This is a targeted recovery for loop- and guard-shaped triggers, not a general fallback; a candidate that satisfies neither Layer~1 nor Layer~1b fails the gate.

\paragraph{Layer~4: Contextual consistency (gate).}
The second gate asks whether the candidate is coherent with the causal chain that produced it: it must sit in the control structure the agent claimed and be reachable from the original critical variable via the recorded alias memory and call chain. A plausible sink in an unreachable branch, or one whose call chain does not connect back to the patched function, is rejected here. A trigger without a coherent call-chain justification is a co-occurrence, not a report, and AutoTrace refuses to emit it.

\paragraph{Corroborating evidence (Layers~2 and~3).}
Layer~2 asks whether existing bounds or null checks already protect the path to~$\sigma$: a complete guard lowers confidence substantially, a partial guard moderately, and a guard present in the fixed commit but absent in the vulnerable one is strong positive evidence, so candidates behind incomplete guards are preserved rather than rejected. Layer~3 compares the two versions: a path blocked in the fixed commit strongly confirms the candidate, since the patch neutralizes it, while a path that persists in both is only weak negative evidence, because many patches add an upstream guard rather than removing the sink, so it lowers confidence modestly without rejecting.

\paragraph{Evidence synthesis: the verification agent.}
Each layer emits a structured packet with a boolean outcome, a confidence, and, where applicable, a witness or guard descriptor. The report predicate is
\[
\mathrm{report}(\sigma) = \mathrm{Gate}_{1/1b}(\sigma)\,\land\, \mathrm{Gate}_{4}(\sigma)\,\land\, \mathrm{Agent3Accept}(\sigma),
\]
a strict \emph{monotone restriction} that fixes what Agent~3 is for: the hard gates are necessary, so a candidate that fails $\mathrm{Gate}_{1/1b}$ or $\mathrm{Gate}_{4}$ is rejected regardless of the agent's confidence, and Agent~3 can only further reject, never admit a gate-failer. It is thus neither an override nor redundant: among gate-passers it weighs the corroborating Layer~2 and Layer~3 signals to render the final verdict, and is the primary decider only for the leak family (CWE-401), whose missing-release defect has no provable forward dataflow (\S\ref{subsec:threats}). It defaults conservatively and can never create a sink, alter the statement, or read a missing witness as success; exact rules and prompts are in Appendix~\ref{app:repro}.

\subsection{Stage~5: SinkTrace-Bench Construction}
\label{subsec:sinktrace-construction}

\paragraph{From triggers to a benchmark.}
Existing vulnerability datasets label \emph{what} is vulnerable but leave \emph{why} implicit, so models learn the surface shape of vulnerable code, the wrong signal for interprocedural triggers whose surface shape is ordinary. SinkTrace-Bench bakes the S2S chain into every sample, matching the supervisory signal to the reasoning task.

\begin{figure*}[t]
\centering
\includegraphics[width=0.92\textwidth]{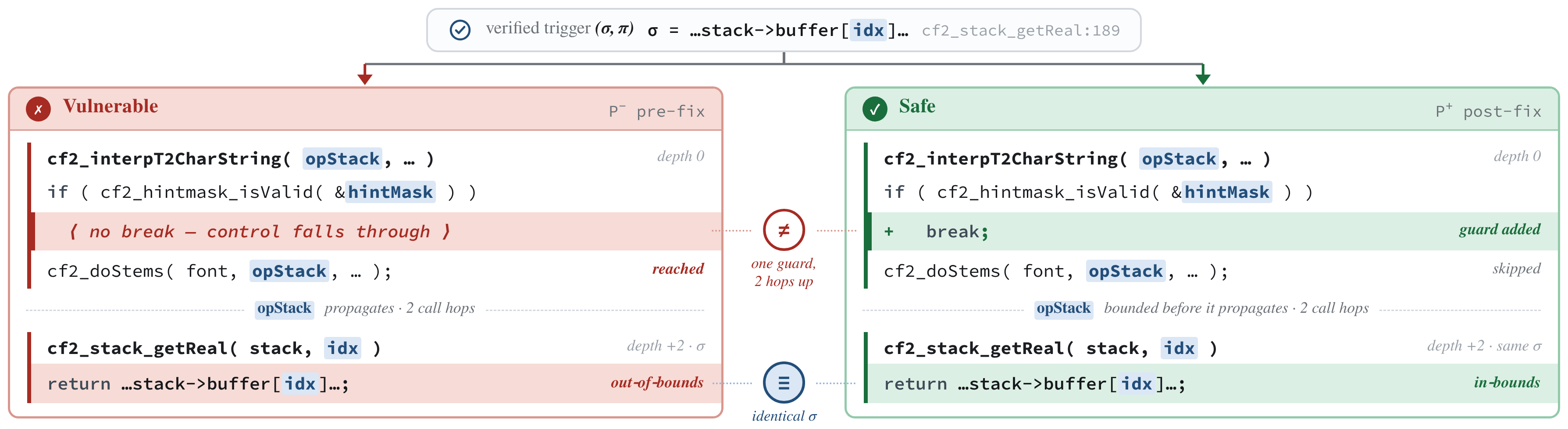}
\caption{Why SinkTrace-Bench is hard (CVE-2014-9659; cf.\ Fig.~\ref{fig:motivation}). Each verifier-confirmed trigger~$(\sigma,\pi)$ (Fig.~\ref{fig:framework}) forks into a matched pair of independently-labeled samples: a \emph{vulnerable} half from the pre-fix snapshot~$P^{-}$ and a \emph{safe} half from the post-fix snapshot~$P^{+}$. Both halves expose the \emph{identical} operative statement~$\sigma$ (\code{stack->buffer[idx]} at \code{cf2\_stack\_getReal:189}, marked~$\equiv$), so a model that keys on the sink alone cannot separate them. The sole difference (marked~$\neq$) is a guard two call hops upstream (the post-fix \code{break} in \code{cf2\_interpT2CharString} at depth~$0$) that bounds \code{opStack} before it propagates to the sink, flipping the label from vulnerable to safe. Assigning these opposite labels therefore demands interprocedural reasoning along the source-to-sink chain rather than patch-diff pattern matching.}
\label{fig:sinktrace_sample}
\end{figure*}

For each accepted trigger the builder emits a \emph{vulnerable} and a \emph{safe} sample (Fig.~\ref{fig:sinktrace_sample}), each carrying the full source-to-sink slice, the trigger statement~$\sigma$, and the critical variable, the vulnerable half from $P^{-}$ and the safe half from $P^{+}$. Because the two come from genuinely different program states, the construction avoids the near-identical-snapshot leakage of patch-only datasets. The labels come from AutoTrace's verifier rather than independent experts, so we call them \emph{verifier-confirmed} and audit their agreement with expert ground truth in \S\ref{subsec:rq3}.

\paragraph{Composition.}
SinkTrace-Bench releases 771 matched vulnerable/safe pairs (1{,}542 samples, exactly balanced) over 540 CVEs, 45 projects, and 16 CWE classes, led by out-of-bounds access (CWE-119/125/787, 55.6\%). We keep only \emph{matched} pairs: a trigger contributes samples only when both its vulnerable ($P^{-}$) and safe ($P^{+}$) half yield a fix-distinguishing slice, and we discard any unpaired half and any byte-identical pair (a residual label leak where the fix falls outside the extracted window), so balance comes from construction and neither class priors nor memorizing ``this CVE is vulnerable'' can separate a pair (\S\ref{subsec:threats}). The two halves are near-identical in length (median 89 vs.\ 91, mean 155 vs.\ 158 lines) and differ by a median of only three changed lines, forcing the decision onto the semantic edit rather than gross textual size. Slices span up to 9 function frames, and 22.7\% are interprocedural.

\paragraph{Deployment to coding agents.}
Because a reported trigger carries a machine-checkable witness, we expose AutoTrace as a Model Context Protocol server and a VSCode extension, so agentic assistants (Claude Code, Codex, Copilot) can call trigger localization as a tool and receive a witness-backed statement rather than an unverified suggestion.

\section{Results and Evaluation}
\label{sec:evaluation}

\noindent\textbf{Research questions.}
We investigate three questions.
RQ1 asks how effective AutoTrace is at trigger localization relative to prior tools on both intra- and interprocedural CVEs.
RQ2 asks which pipeline components contribute to that effectiveness and what the operational cost looks like.
RQ3 asks whether SinkTrace-Bench serves as a discriminative benchmark for interprocedural vulnerability reasoning.

\subsection{Experimental Setup}
\label{subsec:setup}

\paragraph{Dataset.}
We evaluate AutoTrace on InterPVD~\cite{li2024effectiveness}, whose 744 unique real-world C/C++ CVEs span 16~CWE classes; because AutoTrace analyzes each fixing commit as a whole, one sample is one CVE.
InterPVD provides ground-truth annotations for (i) patch statements, (ii) vulnerability-triggering statements and their enclosing functions, and (iii) whether the trigger is interprocedural relative to the patched function; to our knowledge, it is the most comprehensive public benchmark for VFC-conditioned trigger localization.

\paragraph{Metrics.}
We report trigger localization using three metrics.
\emph{VulnHit} counts a CVE as correct only when an identified trigger lands on a ground-truth trigger by matching its statement or line; reaching the correct function without the correct statement earns no credit, matching the line-level criterion VulTrigger~\cite{li2024effectiveness} reports.
\emph{FuncHit} is a broader function-reach metric: it counts a prediction as correct when the predicted function matches either the annotated trigger function or the VFC patch function associated with the benchmark entry, regardless of line.
\emph{Within$\pm$5} is the tolerance-based counterpart to VulnHit: it counts a CVE as correct when any identified trigger lies within a $\pm 5$-line tolerance of a ground-truth trigger line in the correct file or matches a ground-truth trigger statement.
Together, these metrics cover CVE-level trigger success, broad function-level reach, and a tolerance-based localization view.
All three score the trigger \emph{statement}~$\sigma$ and its line and function projections (\S\ref{subsec:problem-definition}), never the witness path: localization is statement-level throughout, and the S2S trace serves only as the certificate behind each scored statement.

\paragraph{Implementation and evaluation cohort.}
All experiments run AutoTrace against InterPVD using the pre-fix snapshot for trigger localization and the post-fix snapshot for differential verification.
The LLM backbone is Gemini-3-Flash; Joern~\cite{joern} constructs CPGs under a pinned repository configuration.
We evaluate over the full InterPVD corpus and count every non-output as a miss: any sample that emits no verified trigger receives a score of 0 on all metrics, so the denominator is fixed at the full corpus and the numbers stay directly comparable to published baselines. Samples that yield no trigger, an acknowledged limitation of the pipeline (\S\ref{subsec:threats}), are already absorbed as misses in every score reported here. RQ3 evaluates SinkTrace-Bench on its own dataset.

\subsection{RQ1: Effectiveness on InterPVD}
\label{subsec:rq1}

\paragraph{RQ1.1: Overall effectiveness vs.\ prior tools.}
Table~\ref{tab:rq1_main} compares AutoTrace with prior tools and a simple LLM baseline on InterPVD.
AutoTrace achieves 75.0\% VulnHit and 80.8\% FuncHit. VulnHit is a strict CVE-level inclusion criterion: a vulnerability counts as solved only when a reported trigger lands on a ground-truth trigger line or statement, not merely the correct function. This is the same line-level criterion VulTrigger~\cite{li2024effectiveness} reports, so on the same 744-CVE corpus the two are directly comparable: AutoTrace's 75.0\% improves on VulTrigger's 69.8\% by 5.2\,pp. VulTrigger publishes no function-level figure, so FuncHit is AutoTrace-only.
The LLM+RAG baseline exposes what this strict criterion measures. On the 593 CVEs both cover, AutoTrace localizes the exact trigger nearly twice as often (71.0\% vs.\ 37.8\% VulnHit): the baseline names the right function but misses the triggering statement. Its higher FuncHit (87.5\% vs.\ 80.8\%) is a coverage artifact, since a plain LLM always emits some function while AutoTrace's verifier abstains rather than guess.
We isolate the contribution of agentic exploration, against a non-agentic baseline sharing the same verifier, in \S\ref{subsec:rq2}.

\begin{table}[!t]
\centering
\footnotesize
\setlength{\tabcolsep}{4pt}
\renewcommand{\arraystretch}{1.05}
\caption{Trigger-localization on InterPVD.}
\label{tab:rq1_main}
\begin{tabularx}{\columnwidth}{@{} >{\raggedright\arraybackslash}X
  S[table-format=2.1] S[table-format=2.1] @{}}
\toprule
\rowcolor{tblhead}
\textbf{Method} & \multicolumn{2}{c}{\textbf{Score (\%)}} \\
\midrule
\rowcolor{tblsub}
\multicolumn{3}{@{}l}{\small\textbf{Static analysis:} \textit{trigger-overlap accuracy}} \\[-2pt]
Flawfinder\,\cite{li2024effectiveness}           & \multicolumn{2}{c}{9.8}  \\
Checkmarx\,\cite{li2024effectiveness}            & \multicolumn{2}{c}{12.7} \\
Infer$^{\dag}$\,\cite{fbinfer2023}               & \multicolumn{2}{c}{2.5}  \\
Fortify$^{\dag}$\,\cite{fortify2023}             & \multicolumn{2}{c}{13.0} \\
CodeQL$^{\dag}$\,\cite{li2024effectiveness}      & \multicolumn{2}{c}{8.6}  \\
\midrule
\rowcolor{tblsub}
\multicolumn{1}{@{}l}{\small\textbf{DL detectors:} \textit{triggering-fn.\ accuracy}}
  & {\small\textit{Intra}} & {\small\textit{Inter}} \\[-2pt]
VulBERTa\,\cite{li2024effectiveness}             & 36.5 & 31.9 \\
LineVul\,\cite{fu2022linevul}                    & 39.3 & 28.1 \\
Devign\,\cite{zhou2019devign}                    & 35.5 & 34.0 \\
ReVeal\,\cite{chakraborty2022reveal}             & 36.6 & 31.5 \\
VulCNN\,\cite{wu2022vulcnn}                      & 35.9 & 36.2 \\
\midrule
\rowcolor{tblsub}
\multicolumn{1}{@{}l}{\small\textbf{Trigger localization}}
  & {\small\textit{VulnHit}} & {\small\textit{FuncHit}} \\[-2pt]
VulTrigger\,\cite{li2024effectiveness}$^{\ddag}$ & 69.8 & {\small n/a} \\
LLM\,+\,RAG$^{\S}$                              & 37.8 & 87.5 \\
\rowcolor{tblhl}
\textbf{AutoTrace (ours)}                        & \bfseries 75.0 & \bfseries 80.8 \\
\bottomrule
\end{tabularx}
{\scriptsize\raggedright
$^{\dag}$72-CVE Magma subset; requires compilation.
$^{\ddag}$VulTrigger reports the same line-level inclusion accuracy as VulnHit; it provides no function-level figure (n/a).
$^{\S}$Retrieval-augmented Qwen3.5 on the 593 InterPVD CVEs with available source.\par}
\end{table}

Three observations follow. Static rule tools cluster at or below 13\% across all five tools (2.5\% to 13.0\%), because their enumerated sink lists cannot cover the open-ended set of project-specific wrappers and kernel helpers that real C/C++ codebases use; Fortify and Infer, which require compilation, perform similarly to open-source scanners on the 72-CVE Magma subset. Every deep-learning detector degrades on interprocedural samples: LineVul's 11.2\,pp drop (39.3\% to 28.1\%) is the largest, confirming that attention-based line ranking cannot follow a value across a call boundary, while VulCNN is nearly flat (35.9\% vs.\ 36.2\%), likely because its image-based representation encodes less positional context to lose. A patch-function classifier~\cite{li2024effectiveness} achieves only 8.4\% F1 on inter-procedural triggers, confirming that detecting patch scope is not the same as locating the operative statement.

\paragraph{RQ1.2: Effectiveness across vulnerability characteristics.}
Table~\ref{tab:rq1_char} breaks the full benchmark down by depth, CWE family, and project. VulnHit degrades with depth (78.0\% at depth~1 to 57--59\% at depth~3 and beyond), and FuncHit tracks it closely throughout, sitting a roughly uniform 5--7\,pp above.

\emph{The function-to-statement gap.}
FuncHit exceeds VulnHit by a roughly uniform 5--7\,pp at every depth (5.8\,pp overall: 80.8\% vs.\ 75.0\%). The gap is the \emph{patch-site fallback}: cases landing in the correct function but on a different statement than InterPVD annotates, because the verifier reports the patched statement as a proxy when an alias, macro wrapper, or missing callee body blocks certification. Since the gap does not widen with depth, cross-boundary failures miss the function outright rather than mislocate within it; in every regime the verifier turns uncertainty into abstention, not false positives.

\begin{table}[!t]
\centering
\footnotesize
\setlength{\tabcolsep}{4pt}
\renewcommand{\arraystretch}{1.03}
\caption{AutoTrace breakdown by interprocedural depth, CWE family, and project (744 CVEs).}
\label{tab:rq1_char}
\begin{tabularx}{\columnwidth}{@{}>{\raggedright\arraybackslash}X S[table-format=3.0] S[table-format=2.1] S[table-format=2.1] S[table-format=2.1]@{}}
\toprule
\rowcolor{tblhead}
\textbf{Subset} & {\textbf{N}} & {\textbf{VulnHit}} & {\textbf{FuncHit}} & {\textbf{W$\pm$5}} \\
\midrule
\rowcolor{tblsub}
\multicolumn{5}{@{}l}{\small\textbf{Call depth from patched function}} \\[-2pt]
Depth~1 (intra)   & 564 & 78.0 & 83.9 & 80.1 \\
Depth~2           &  98 & 72.4 & 77.6 & 73.5 \\
Depth~3           &  53 & 56.6 & 62.3 & 58.5 \\
Depth~$\geq\!4$   &  29 & 58.6 & 65.5 & 65.5 \\
\midrule
\rowcolor{tblsub}
\multicolumn{5}{@{}l}{\small\textbf{CWE family}} \\[-2pt]
Memory \tiny(CWE-119/125/787)     & 413 & 75.3 & 81.1 & 77.2 \\
Integer \tiny(CWE-189/190/191)    & 106 & 73.6 & 76.4 & 73.6 \\
NULL ptr \tiny(CWE-476)           &  82 & 68.3 & 75.6 & 73.2 \\
Resource leak \tiny(CWE-401/772)  &  40 & 80.0 & 95.0 & 87.5 \\
Use-after-free \tiny(CWE-416)     &  31 & 71.0 & 77.4 & 71.0 \\
\midrule
\rowcolor{tblsub}
\multicolumn{5}{@{}l}{\small\textbf{Project} \textit{(top by N)}} \\[-2pt]
\texttt{linux}       & 251 & 70.9 & 73.3 & 72.1 \\
\texttt{ffmpeg}      & 102 & 80.4 & 82.4 & 80.4 \\
\texttt{tcpdump}     &  54 & 72.2 & 87.0 & 77.8 \\
\texttt{imagemagick} &  54 & 66.7 & 81.5 & 72.2 \\
\texttt{qemu}        &  49 & 77.6 & 87.8 & 81.6 \\
\midrule
\rowcolor{tblhl}
\textbf{Total}       & \textbf{744} & \textbf{75.0} & \textbf{80.8} & \textbf{77.2} \\
\bottomrule
\end{tabularx}
\end{table}

\emph{CWE misprediction} is architecturally coherent rather than random: the most frequent confusions are CWE-119$\to$125, 119$\to$129, 772$\to$401, and 119$\to$787, in each case between classes that share the same structural trigger pattern, so downstream slicing proceeds on a plausible target even when the label is imprecise.

\paragraph{RQ1.3: Output richness.}
AutoTrace emits 2.15 deduplicated triggers per CVE, between expert ground truth (1.20) and VulTrigger (2.30): more than experts because most CVEs expose several critical variables, fewer than VulTrigger because every trigger needs a CPG-confirmed path.

\subsection{RQ2: Component Contribution and Operational Cost}
\label{subsec:rq2}

\paragraph{RQ2.1: Component contribution.}
\emph{Stage~1 (CV extraction).} Stage~1 reaches 43.2\% strict CWE accuracy, but most misclassifications are hierarchically valid (child, parent, or root-cause relations) and fuzzy CV recall reaches 82.7\%. Classification quality matters downstream: a correct predicted CWE yields a 59.5\% trigger-found rate versus 52.1\% when wrong, so CWE context acts as a search-space constraint, not just a label. On 337 commits with curated single-hunk ground truth, Stage~1 picks a critical variable from the correct hunk in 97.9\% of cases. This matters because VulTrigger requires the security-relevant diff hunk to be supplied manually, whereas AutoTrace must autonomously identify it from commits that often bundle refactoring with the security fix; the 2.1\% failures are multi-hunk commits where a high-churn change displaced the true fix.

\emph{Verification pipeline.} A candidate enters the output only through the layered gates of \S\ref{subsec:verification}. Most candidates carry a direct dataflow witness and are admitted without the synthesis agent; only the 360 ambiguous candidates reach Agent~3, which accepts 356 (98.9\%). Agent~3 therefore tightens acceptance at the margin rather than driving it, and Table~\ref{tab:layer_contrib} reports the evidence mix for these 360.

\emph{Per-layer contribution.} Among the 360 candidates that reach Agent~3 (Table~\ref{tab:layer_contrib}), 55.6\% carry a direct L1 dataflow witness, 33.1\% rely on the L1b loop-bound/control-dependency fallback that recovers otherwise-failing candidates, and 11.4\% are leak-family admissions with L1 skipped; no complete-L1-failure candidate reaches the agent, since hard-gate rejections precede it. Among these, L2 flagged a protective guard in 60.8\% (never rejecting) and L3 confirmed \code{flow\_blocked} in 64.7\% (its strongest positive signal), while all 360 are L4-unguarded. In short, L1/L4 supply hard admissibility, L1b recovers a third of otherwise-failing candidates, and L2/L3 only calibrate confidence.

\begin{table}[!t]
\centering\footnotesize\setlength{\tabcolsep}{4pt}\renewcommand{\arraystretch}{1.05}
\caption{Verification gate activation that reach Agent~3.}
\label{tab:layer_contrib}
\begin{tabularx}{\columnwidth}{@{} l >{\raggedright\arraybackslash}X r @{}}
\toprule
\rowcolor{tblhead}
\textbf{Gate} & \textbf{Signal} & \textbf{\%} \\
\midrule
\rowcolor{tblsub}
\multicolumn{3}{@{}l}{\small\textbf{Admissibility (hard):} \textit{candidate rejected if failed}} \\[-2pt]
L1 direct     & Direct S2S witness to sink        & 55.6 \\
L1b control   & Loop/guard fallback (recovered)   & 33.1 \\
L1 bypass     & Leak-family CWE; L1 waived        & 11.4 \\
L4 context    & Sink reachable and guard-free     & 100.0 \\
\midrule
\rowcolor{tblsub}
\multicolumn{3}{@{}l}{\small\textbf{Corroboration (soft):} \textit{never rejects}} \\[-2pt]
L2 guard      & Protective guard on path          & 60.8 \\
L3 differential & Patch neutralizes flow          & 64.7 \\
\bottomrule
\end{tabularx}
\end{table}

\emph{Leave-one-out ablation.} On a 25-CVE intraprocedural subset (full system: 100\% FuncHit, 56\% VulnHit), reflection-with-retry contributes most ($-$32\,pp FuncHit), then episodic memory ($-$16), CV refinement ($-$12), and long-term memory ($-$8); disabling all four drops FuncHit to 60\%, so the mechanisms are roughly additive.

\emph{Agentic vs.\ non-agentic exploration.} The ablation above varies components \emph{within} the agent; a sharper question is what agentic exploration adds over ordinary interprocedural traversal. We run a non-agentic baseline that shares AutoTrace's CPG, slicing, sink taxonomy, and verifier gates but replaces the LLM exploration agent with a deterministic best-first traversal over call edges, on the depth-2+ cohort where the question is sharpest. On the 43~CVEs both arms complete, the agentic explorer raises VulnHit from 4.7\% to 18.6\% (and at depth~2 from 7.7\% to 30.8\%), while FuncHit stays comparable (69.8\% vs.\ 74.4\%): its contribution concentrates in VulnHit at depth, where deterministic traversal reaches the right function but cannot track the aliasing and indirect dispatch needed to certify the precise statement. Because the deterministic arm uses an open-weights verifier backbone, we read the VulnHit gap as directional.

\paragraph{RQ2.2: Operational cost profile.}
At \$0.65 per CVE on average (Gemini-3-Flash billing) and 23.2 LLM calls per CVE, AutoTrace is inexpensive in model cost; the wall-clock is instead dominated by Joern CPG construction and slicing, not LLM inference, with a median of 42~minutes and a P90 tail near 13.8~hours on deep CVEs. Because the dominant cost is per repository, a deployment localizing many CVEs against one codebase builds the graph once and amortizes it, and the low token cost makes corpus-scale SinkTrace-Bench generation economical. AutoTrace therefore targets offline batch localization, not interactive use.

\subsection{RQ3: SinkTrace-Bench as a Discriminative Benchmark}
\label{subsec:rq3} 
Beyond training data, SinkTrace-Bench is a point-in-time evaluation benchmark for interprocedural reasoning. Each sample presents one vulnerable or patched program state and asks a model to classify it as \emph{Vulnerable} or \emph{Safe}. This is distinct from the localization of \S\ref{subsec:rq1}: the $(\sigma,\pi)$ trace is the input a model reasons over, not an output to be localized.

\emph{Prompt inputs and contamination control.} To test reasoning rather than recall, every model receives the CWE category, the critical variable, and the source-to-sink slice (vulnerable or patched), and outputs a binary \texttt{decision} plus trigger localization (function, line, statement) when vulnerable. No CVE identifier, file name, or commit hash is shown, and inline \code{VULNERABLE}/\code{FIX}/\code{SAFE} markers are stripped; the CWE and critical variable focus reasoning on the relevant path rather than supply a retrieval key, and cannot alone recover the CVE.

\begin{figure}[!t]
\centering
\includegraphics[width=\columnwidth]{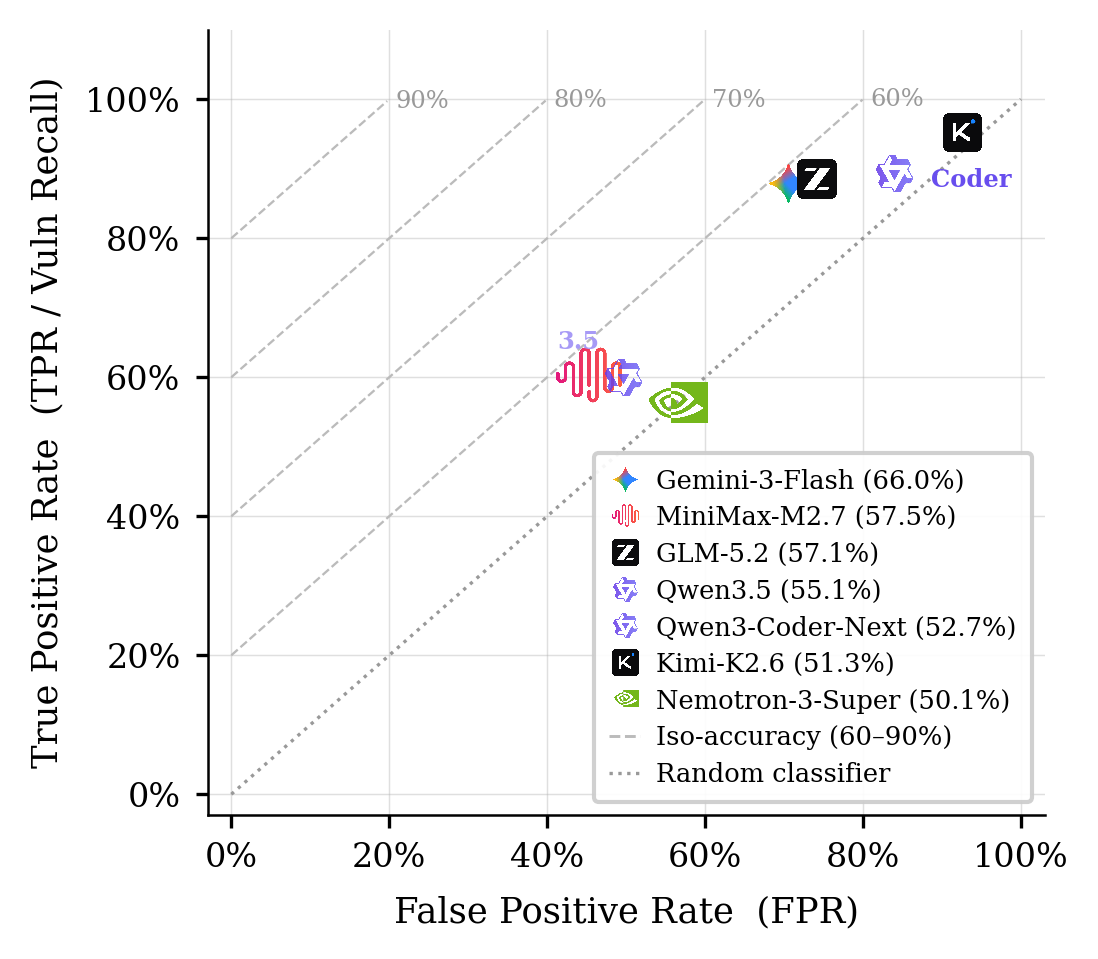}
\caption{Zero-shot ROC comparison of seven frontier models on SinkTrace-Bench.}
\label{fig:sinktrace_eval}
\end{figure}

\emph{Results.} Figure~\ref{fig:sinktrace_eval} and Table~\ref{tab:sinktrace_results} report zero-shot performance on the 1{,}542-sample matched-pair release; because the set is balanced, Accuracy equals macro-F1 at full coverage. All seven backbones (MiniMax-M2.7, GLM-5.2, Qwen3.5, Qwen3-Coder-Next, Kimi-K2.6, Nemotron-3-Super, Gemini-3-Flash) are evaluated over the whole set. No model clears the benchmark: the top accuracy is 59.0\% (Gemini-3-Flash), only nine points above the 50\% chance floor, and it reaches even that by defaulting to \emph{vulnerable}. The dominant failure mode is a true-positive/true-negative asymmetry: the strongest backbones (Gemini-3-Flash, GLM-5.2, Qwen3-Coder-Next, Kimi-K2.6) reach 87--95\% TPR but collapse on the safe half (7--31\% TNR), predicting a \emph{vulnerable} verdict whenever the sink is present; in ROC space (Fig.~\ref{fig:sinktrace_eval}) they land at high FPR while most models hug the random-classifier diagonal. The one model that stays balanced is the reasoning-enabled MiniMax (54.7\% TNR against 60.3\% TPR), which leads on both TNR and FPR though not on raw accuracy, indicating that separating a matched pair requires reasoning about the upstream patch rather than pattern-matching the sink. This also argues against contamination as the dominant factor: a model recalling ``this CVE is vulnerable'' would not explain the systematic collapse on the \emph{safe} twins, and the spread tracks code-reasoning ability rather than mere exposure.

\emph{Label fidelity.} Because the labels come from AutoTrace's verifier rather than independent experts, the benchmark carries a circularity risk bounded by the verifier's reliability. Auditing against the independent InterPVD annotations, for the vulnerable samples whose CVE has an expert trigger the SinkTrace label agrees with the expert function in 64.8\% of cases and the exact line in 28.0\%. This mirrors AutoTrace's own localization accuracy, so the labels inherit rather than exceed the localizer's reliability; we call them verifier-confirmed, with the function-level label the more reliable signal.

\subsection{Threats to Validity}
\label{subsec:threats}
\paragraph{Non-completions.}
Of the 744 CVEs, 603 (81\%) yield at least one verified trigger; the remaining 141 produce none and are counted as misses in the fixed denominator (\S\ref{subsec:setup}) rather than behind a smaller cohort, so the headline metrics already absorb the cost. The dominant cause is a 3-hour wall-time timeout on deep multi-file vulnerabilities, where Joern CPG construction or slicing dominates cost; the remainder are API or tooling failures and incomplete runs. Because the timeouts skew deep, the depth-3+ figures (Table~\ref{tab:rq1_char}) are the conservative reading of the hardest tail. Building the CPG over the full call path (\S\ref{sec:discussion}), relaxing the wall-time limit, and improving CPG incremental loading are the practical mitigations.
\begin{table}[!t]
  \centering
  \footnotesize\setlength{\tabcolsep}{4pt}\renewcommand{\arraystretch}{1.05}
  \caption{Zero-shot frontier-model results on SinkTrace-Bench (all 1{,}542 samples; values \%). \textbf{Bold}\,=\,best per column.}
  \label{tab:sinktrace_results}
  \begin{tabularx}{\columnwidth}{@{} >{\raggedright\arraybackslash}X S[table-format=2.1] S[table-format=2.1] S[table-format=2.1] S[table-format=2.1] S[table-format=2.1] S[table-format=2.1] @{}}
    \toprule
    \textbf{Model} & \textbf{Acc} & \textbf{F1} & \textbf{TPR} & \textbf{TNR} & \textbf{FPR} & \textbf{FNR} \\
    \midrule
    Gemini\mbox{-}3\mbox{-}Flash    & \bfseries 59.0 & \bfseries 68.1 & 87.5 & 30.5 & 69.5 & 12.5 \\
    MiniMax\mbox{-}M2.7             & 57.5 & 58.7 & 60.3 & \bfseries 54.7 & \bfseries 45.3 & 39.7 \\
    GLM\mbox{-}5.2                  & 57.1 & 67.4 & 88.5 & 25.8 & 74.2 & 11.5 \\
    Qwen3.5\mbox{-}9B               & 55.1 & 57.2 & 59.9 & 50.3 & 49.7 & 40.1 \\
    Qwen3\mbox{-}Coder\mbox{-}Next  & 52.7 & 65.3 & 89.2 & 16.1 & 83.9 & 10.8 \\
    Kimi\mbox{-}K2.6                & 51.3 & 66.2 & \bfseries 95.2 & 7.4 & 92.6 & \bfseries 4.8 \\
    Nemotron\mbox{-}3\mbox{-}Super  & 50.1 & 53.0 & 56.3 & 43.8 & 56.2 & 43.7 \\
    \bottomrule
  \end{tabularx}
\end{table}
\paragraph{Internal.}
Three internal limits stem from the CPG substrate. Resource-leak CWEs (401 and kin) manifest as a \emph{missing} deallocation, which dataflow cannot prove, so the verifier admits them on structural grounds, widening the hallucination surface the dataflow gate ordinarily closes (all such decisions are logged). Joern's field-sensitivity does not fully track nested dereferences, so slices for deeply indirect pointer patterns may omit use sites; because the verifier reasons over the same graph, this is an integrity limit as well as a recall one (\S\ref{subsec:threat-model}). More often the loss is coarser: on 80 CVEs the patch-scoped CPG (\S\ref{subsec:slicing}) never grows to the sink function or the sink hides behind an unexpanded macro, so Joern returns an \emph{empty slice}, not from an inability to follow the value but from being handed a partial graph (remedy in \S\ref{sec:discussion}). And InterPVD does not uniformly annotate root-cause triggers, so a causally correct trigger can score as a VulnHit miss, the bias \S\ref{sec:discussion} quantifies.

\paragraph{External.}
InterPVD covers C/C++ CVEs over 16~CWE classes; generalization to managed languages, firmware, or fuzz-discovered classes is unvalidated. The subtler threat is contamination: the CVEs predate the evaluated models' cutoffs, so a model could recall a vulnerability rather than reason about it, inflating RQ3 and, to a lesser degree, RQ1. We reduce but cannot eliminate it (RQ3 prompts omit CVE identifiers, file names, and commit hashes (\S\ref{subsec:rq3}); matched vulnerable/patched pairs make ``this CVE is vulnerable'' non-discriminative; and model-to-model variation under identical inputs checks against pure recall), but a definitive control requires vulnerabilities disclosed after every model's cutoff, which the current corpus cannot supply. This motivates our main forward direction: running AutoTrace over repository snapshots dated after every evaluated model's cutoff, and ultimately over unpatched code where no fix or advisory yet exists, so no public record of the trigger predates the model and the memorization route is removed rather than merely bounded. Because each reported trigger is CPG-gated rather than a model guess, the same leak-free setting doubles as zero-day trigger discovery: evidence-backed candidates in live code rather than the reproduction of known CVEs. RQ1 is less exposed in any case, demanding an exact statement on a specific snapshot.

\paragraph{Construct.}
VulnHit credits a trigger that lands on the ground-truth statement or line and FuncHit ignores line position, with Within$\pm$5 between; the three are meant to be read together. AutoTrace's accuracy is also bounded by its backbone: a preliminary small-backbone run produced no verified triggers, so the pipeline presupposes a frontier-scale model. Finally, matched-pair construction retains a CVE only when both halves yield a resolvable, fix-distinguishing slice, biasing the benchmark toward cleanly sliceable vulnerabilities.

\paragraph{Benchmark construction and manual audit.}
The pipeline emitted 2{,}801 candidate samples from the InterPVD VFCs, curated in three steps: drop candidates whose source file Joern cannot reconstruct; remove exact duplicates, since a CVE's several critical variables share one \emph{source-to-sink} chain and collapse to identical (slice, trigger, label) triples; and keep only \emph{matched} vulnerable/safe pairs whose halves differ, discarding unpaired halves and byte-identical pairs (residual label leaks where the fix falls outside the extracted window). Each slice keeps every frame on the source-to-sink path so the critical variable's provenance stays continuous, with frames beyond a window cropped around their next-hop call site. Earlier snapshots (2{,}414 body-only, 1{,}687 endpoint-only, 1{,}556 pre-full-path) predate this representation; Table~\ref{tab:sinktrace_results} and the audit below report the current release. To meet the reviewers' request for an independent check, we manually annotated all 1{,}542 samples against a fixed eight-field rubric. Every sample carries an agreed Vulnerable/Safe label (100\% label fidelity: 92.3\% agree outright, 7.7\% agree after deliberation), and the CWE family, source-to-sink validity, and safe-side fix are consistent in every sample. The one imperfection is trigger precision: the annotated trigger statement genuinely causes the vulnerability in 83.8\% of vulnerable samples (646/771, with 6 further partial matches), still well above the 64.8\% automated agreement with InterPVD's expert function (\S\ref{subsec:rq3}) because annotators credit root-cause triggers that diverge from InterPVD's consequence-level annotation; the remaining 15.4\% point off-target where Joern cannot resolve a cross-function hop (alias, indirect call, or macro wrapper). We release every per-sample annotation and position SinkTrace-Bench for training and reasoning evaluation rather than as a gold detection oracle.

\section{Discussion}
\label{sec:discussion}

Two findings shape how AutoTrace should be read. The 5.8\,pp FuncHit/VulnHit gap reflects conservative abstention, not confident error: at every depth the system reaches the correct function but reports a proxy statement when it cannot certify the exact operation (\S\ref{subsec:rq1}). The strictness of the dataflow gate is the intended price of grounded acceptance, so the principal lever for recall is CPG completeness, not verifier relaxation (fully graphed \code{radare2} 100\% FuncHit vs.\ deep, timeout-prone \code{linux} 73.3\%), and at \$0.65 per CVE the same machinery yields SinkTrace-Bench at a scale no prior localizer reaches without manual annotation.

\emph{From patch-scoped to call-path CPGs.} The completeness gap above is an engineering limit, not a reasoning one, and closing it is a central near-term direction. AutoTrace grows the workspace on demand from the patch (\S\ref{subsec:slicing}) to bound context and whole-repository noise, but that growth stalls at macro-hidden or unresolved call targets, so the sink function is often absent and the slice comes back empty. Building the CPG eagerly over every file on the source-to-sink call path, and expanding macros and headers so macro-sinks (e.g.\ \code{ipaddr\_string}) become real nodes, would let Joern's dataflow carry the critical variable through pointers, struct fields, and calls uniformly rather than re-anchoring it by name at each hop. Field-sensitivity still misses some deep aliasing (\S\ref{subsec:threats}), so completeness stays below 100\%, but this moves AutoTrace from cleanly sliceable cases to the large majority, handled the same way across scenarios.

\section{Conclusion}
\label{sec:conclusion}

Trigger localization demands causal, interprocedural reasoning that neither static rule sets nor attention-based line rankers provide. AutoTrace supplies it with an agentic CPG search whose every reported trigger is gated by deterministic admissibility, reaching 75.0\% VulnHit and 80.8\% FuncHit on the full InterPVD benchmark and improving on VulTrigger's 69.8\% under the same inclusion metric; the same machinery yields SinkTrace-Bench (1{,}542 verifier-confirmed samples, 771 matched pairs, at \$0.65 per CVE). Our primary next step is to run AutoTrace over live repository snapshots to surface zero-day triggers, where no prior public record can leak the answer and its CPG-gated acceptance keeps discovered candidates evidence-backed; extending to managed languages and distilling localization into smaller models follow.

\let\oldthebibliography\thebibliography
\renewcommand{\thebibliography}[1]{%
  \oldthebibliography{#1}%
  \setlength{\itemsep}{-0.4ex}%
  \setlength{\parsep}{0pt}%
}
\bibliographystyle{IEEEtran}
\IfFileExists{paper/refs.bib}{%
  \bibliography{paper/refs}%
}{%
  \bibliography{refs}%
}

%% ── Joern query box: light orange ──────────────────────────────
\definecolor{joernbg}{RGB}{255,245,228}
\definecolor{joernframe}{RGB}{185,90,18}
\tcbset{joernbox/.style={
  enhanced,breakable,width=\columnwidth,
  colback=joernbg,colframe=joernframe,
  arc=2pt,outer arc=2pt,boxrule=0.6pt,
  before skip=1pt, after skip=1pt,
  left=3pt,right=3pt,top=1pt,bottom=1pt,
  fonttitle=\sffamily\bfseries\footnotesize,
  colbacktitle=joernframe,coltitle=white,
  titlerule=0pt,toptitle=1pt,bottomtitle=1pt,
}}

\appendix
\section{Reproducibility Supplement}
\label{app:repro}

\subsection{Agent Prompts}
\label{app:prompts}

\begin{tcolorbox}[promptbox, title={Agent~1: Critical-Variable Extraction}]
\begin{lstlisting}[language={},basicstyle=\ttfamily\tiny,breaklines=true,
  aboveskip=0pt,belowskip=0pt,
  numbers=none,frame=none,backgroundcolor=\color{promptbg}]
You are a security expert analyzing code commits. Extract ALL distinct
critical variables from this fix using exact code identifiers (struct
fields: the field, not the pointer); predict a CWE per variable.
In: {commit_message, diff, context}   Out: [{name, location, reason, confidence, predicted_cwe}]
\end{lstlisting}
\end{tcolorbox}

\begin{tcolorbox}[promptbox, title={Agent~2: Frontier Exploration}]
\begin{lstlisting}[language={},basicstyle=\ttfamily\tiny,breaklines=true,
  aboveskip=0pt,belowskip=0pt,
  numbers=none,frame=none,backgroundcolor=\color{promptbg}]
You are a vulnerability analyst tracing a critical variable through C/C++
to find sinks. Tools: slice_variable, find_tainted_callsites, list_callees/
callers, read_function_source_code, lookup_cwe_sink_patterns. Local first; <=5 targets.
In: {cve_id, cwe, function, variable, depth, slice_nodes, ...}   Out: ExplorationDecision {sinks, targets}
\end{lstlisting}
\end{tcolorbox}

\begin{tcolorbox}[promptbox, title={Agent~3: Verification Judge}]
\begin{lstlisting}[language={},basicstyle=\ttfamily\tiny,breaklines=true,
  aboveskip=0pt,belowskip=0pt,
  numbers=none,frame=none,backgroundcolor=\color{promptbg}]
You are a vulnerability verification specialist. Given L1-L4 evidence for one
candidate, decide if it is a TRUE trigger. Rules: L1 dataflow required (skip
for leak CWEs); L2/L3 adjust confidence ("flow_exists_both" never rejects alone); L4 confirms guard absence.
In: {evidence: layer1..layer4}   Out: {is_true_trigger, confidence, reasoning}
\end{lstlisting}
\end{tcolorbox}

\begin{tcolorbox}[promptbox,
  left=3pt,right=3pt,top=1pt,bottom=1pt,
  toptitle=1pt,bottomtitle=1pt,
  title={SinkTrace-Bench Evaluation: Frontier Models (RQ3)}]
\begin{lstlisting}[language={},basicstyle=\ttfamily\tiny,breaklines=true,
  aboveskip=0pt,belowskip=0pt,
  numbers=none,frame=none,backgroundcolor=\color{promptbg}]
[SYSTEM] You are a vulnerability analysis expert. Given a source-to-sink
slice, decide Vulnerable or Safe; if vulnerable, give the exact trigger
statement. Respond ONLY: {"decision":"Vulnerable"|"Safe","trigger_function",
"trigger_line","trigger_statement"}  (localization null when Safe).
[USER] CWE: {cwe}  Critical Variable: {variable}
## Source-to-Sink Code:  {source_to_sink_code}
Return JSON only.
\end{lstlisting}
\end{tcolorbox}

\lstdefinestyle{joernq}{language={},basicstyle=\ttfamily\tiny,breaklines=true,
  aboveskip=0pt,belowskip=0pt,numbers=none,frame=none,
  backgroundcolor=\color{joernbg}}

\begin{tcolorbox}[joernbox, title={Appendix B: Joern (CPGQL) traversals per evidence layer}]
\label{app:queries}
{\scriptsize Notation: $m$~= method, $v$~= critical variable, $L$~= sink line. Shared bindings below.}
\begin{lstlisting}[style=joernq]
val tgt = cpg.identifier.name(v) ++ cpg.fieldIdentifier.canonicalName(v)
val src = m.ast.isIdentifier.nameExact(v) ++ m.parameter.nameExact(v)
\end{lstlisting}
{\scriptsize\emph{slice\_variable}: backward, then forward data-flow slice of $v$.}
\begin{lstlisting}[style=joernq]
tgt.reachableByFlows(m.parameter ++ m.call ++ m.literal)
tgt.reachableByFlows(m.call.argument ++ m.ast.isReturn)
\end{lstlisting}
{\scriptsize\emph{find\_tainted\_callsites}: taint from $v$ to a call argument, mapped to callee/param.}
\begin{lstlisting}[style=joernq]
m.call.argument.reachableByFlows(src).map(_.inCall)
\end{lstlisting}
{\scriptsize\emph{L1 has\_flow}: sink reachable from the source (out-parameter semantics).}
\begin{lstlisting}[style=joernq]
val snk = m.ast.lineNumber(L) ++ m.ast.code(".*" + sink + ".*")
snk.collect { case n: CfgNode => n }.reachableByFlows(src)
\end{lstlisting}
{\scriptsize\emph{L1b control dependency}: a loop header that references $v$.}
\begin{lstlisting}[style=joernq]
m.ast.isControlStructure.filter(_.ast.isIdentifier.nameExact(v).nonEmpty)
\end{lstlisting}
{\scriptsize\emph{L3 differential}: rerun L1 on patched CPG; flow blocked in $P^+$ only.}
\end{tcolorbox}

\end{document}